\documentclass[aps,twocolumn,prl,longbibliography,superscriptaddress]{revtex4-2}
\usepackage[colorlinks=true,allcolors=metalblue]{hyperref}
\usepackage{orcidlink}
\usepackage{amsmath,amsfonts,amssymb,graphicx,bm,color}
\usepackage[T1]{fontenc}
\usepackage{graphicx}
\usepackage{dcolumn}
\usepackage{bm}
\usepackage{color}
\usepackage{amsmath}
\usepackage{amssymb}
\usepackage{hyperref}
\usepackage{subfigure}
\usepackage{float}
\usepackage{bigints}
\usepackage{changes}
\usepackage{braket}

\definecolor{metalblue}{HTML}{397378}

\begin{document}

\author{L. Brodoloni \orcidlink{0009-0002-0887-4020}}
\affiliation{
	School of Science and Technology, Physics Division, Universit\`a di Camerino, 62032 Camerino, Italy}
\affiliation{
	INFN, Sezione di Perugia, I-06123 Perugia, Italy
}
\author{J. Vovrosh \orcidlink{0000-0003-4034-5786}}
\affiliation{PASQAL SAS, 24 rue Emile Baudot - 91120 Palaiseau,  Paris, France}

\author{S. Juli\`a-Farr\'e \orcidlink{0000-0003-4034-5786}}
\affiliation{PASQAL SAS, 24 rue Emile Baudot - 91120 Palaiseau,  Paris, France}

\author{A. Dauphin \orcidlink{0000-0003-4996-2561}}
\affiliation{PASQAL SAS, 24 rue Emile Baudot - 91120 Palaiseau,  Paris, France}

\author{S. Pilati \orcidlink{0000-0002-4845-6299}}
\affiliation{
	School of Science and Technology, Physics Division, Universit\`a di Camerino, 62032 Camerino, Italy}
\affiliation{
	INFN, Sezione di Perugia, I-06123 Perugia, Italy
}

\title{Spin-glass quantum phase transition in amorphous arrays of Rydberg atoms}

\begin{abstract}
The experiments performed with neutral atoms trapped in optical tweezers and coherently coupled to the Rydberg state allow quantum simulations of paradigmatic Hamiltonians for quantum magnetism. Previous studies have focused mainly on periodic arrangements of the optical tweezers, which host various spatially ordered magnetic phases. Here, we perform unbiased quantum Monte Carlo simulations of the ground state of quantum Ising models for amorphous arrays of Rydberg atoms. 
These models are designed to feature well-controlled local structural properties in the absence of long-range order. 
Notably, by determining the Edwards-Anderson order parameter, we find evidence of a quantum phase transition from a paramagnetic to a spin-glass phase. The magnetic structure factor indicates short-range isotropic antiferromagnetic correlations. 
For the feasible sizes, the spin-overlap distribution features a nontrivial structure with two broad peaks and a sizable weight at zero overlap. 
The comparison against results for the clean kagome lattice, which features local structural properties similar to those of our amorphous arrays, highlights the important role of the absence of long-range structural order of the underlying array.
Our findings indicate a route to experimentally implement the details of a Hamiltonian which hosts a quantum spin-glass phase.
\end{abstract}

\maketitle

Frustrated random interactions in Ising models can give rise to spin-glass phases, which exhibit a variety of intriguing phenomena, including disordered magnetism, memory effects, and possibly the breakdown of ergodicity~\cite{RevModPhys.58.801,charbonneau2023spin}. 
These models were originally introduced to describe dilute magnetic alloys and play a central role in the study of combinatorial optimization problems~\cite{Barahona1982}. 
In the quantum version, which is of interest here, they are also studied to assess whether quantum annealers can achieve a quantum advantage~\cite{doi:10.1126/science.1068774,tanaka2017quantum,Hauke_2020}.
However, various fundamental questions are still unanswered, e.g., where replica symmetry breaking (RSB) occurs~\cite{PhysRevB.41.4858,PhysRevB.41.428,Thirumalai_1989,PhysRevB.52.384,PhysRevB.39.11828,PhysRevE.92.042107,PhysRevLett.127.207204} or whether the spin-glass transition survives longitudinal fields~\cite{PhysRevE.96.032112,manai2022almeida,kiss2023exact,PhysRevLett.129.220401}. 
Another open problem is the relation between spin-glass and many-body localized phases~\cite{PhysRevLett.113.200405,PhysRevLett.113.107204,Mossi2017,PhysRevLett.125.260405}.

Today, quantum simulations of paradigmatic quantum Ising Hamiltonians are performed with various experimental systems, including ultracold atomic gases~\cite{bloch2012quantum,lewenstein2012ultracold}, trapped ions~\cite{blatt2012quantum}, superconducting qubits~\cite{houck2012chip}, or neutral atoms precisely positioned using optical tweezers and then coupled to the Rydberg state~\cite{Schauss2018,browaeys2020many,henriet_quantum_2020}.
For the latter platform, various magnetically ordered phases have been theoretically predicted~\cite{PhysRevB.93.104412,PhysRevLett.124.103601,PhysRevB.105.174417,o2023entanglement,PhysRevA.108.053314,PhysRevE.106.034121,10.21468/SciPostPhys.14.1.004,doi10.1073pnas.2015785118} and in some cases experimentally observed~\cite{scholl2021quantum,ebadi2021quantum,chen2023continuous}. Several periodic arrangements of the optical tweezers have been considered, including one-dimensional and two-dimensional (2D) configurations, such as square, triangular, and kagome periodic lattices. 
Incommensurate phases that separate ordered and paramagnetic phases have been studied in single-leg and double-leg chains~\cite{PhysRevLett.105.230403,rader2019floating,PhysRevB.106.165124,PhysRevResearch.7.013215,zhang2025probing}.
Interestingly, memory effects associated with glassy behavior have been predicted to occur in a periodic array at the boundary of differently ordered phases~\cite{PhysRevLett.130.206501}. Yet, this finding has been questioned~\cite{hibatallah2024recurrentneuralnetworkwave}.

Frustration effects may be induced by the lattice geometry and/or by random ferromagnetic and antiferromagnetic interactions. In Rydberg-atom platforms, the interatomic couplings are always antiferromagnetic, but the atomic layout is essentially arbitrary~\cite{doi:10.1126/science.abo6587,PRXQuantum.4.010316,kim2022rydberg}. 
This flexibility has been exploited in experiments addressing combinatorial optimization~\cite{doi:10.1126/science.abo6587,PhysRevA.111.032611,cazals2025}  and machine-learning problems~\cite{PhysRevA.107.042615,kornjavca2024large,PhysRevResearch.5.043117}.
In fact, the couplings can be randomized by arranging the positions of the optical tweezers in disordered arrays~\cite{PhysRevLett.124.130604}. 
As recently shown~\cite{juliafarre2024amorphous}, this feature allows one to create amorphous solids, namely configurations featuring well-defined short-range structural properties in the absence of long-range positional or orientational orders. However, the phase diagram of quantum Ising models with positionally disordered couplings is essentially unknown, and it is unclear whether antiferromagnetic couplings suffice to give rise to spin-glass phases.

\begin{figure}[h]
\centering
\includegraphics[width=1.\columnwidth]{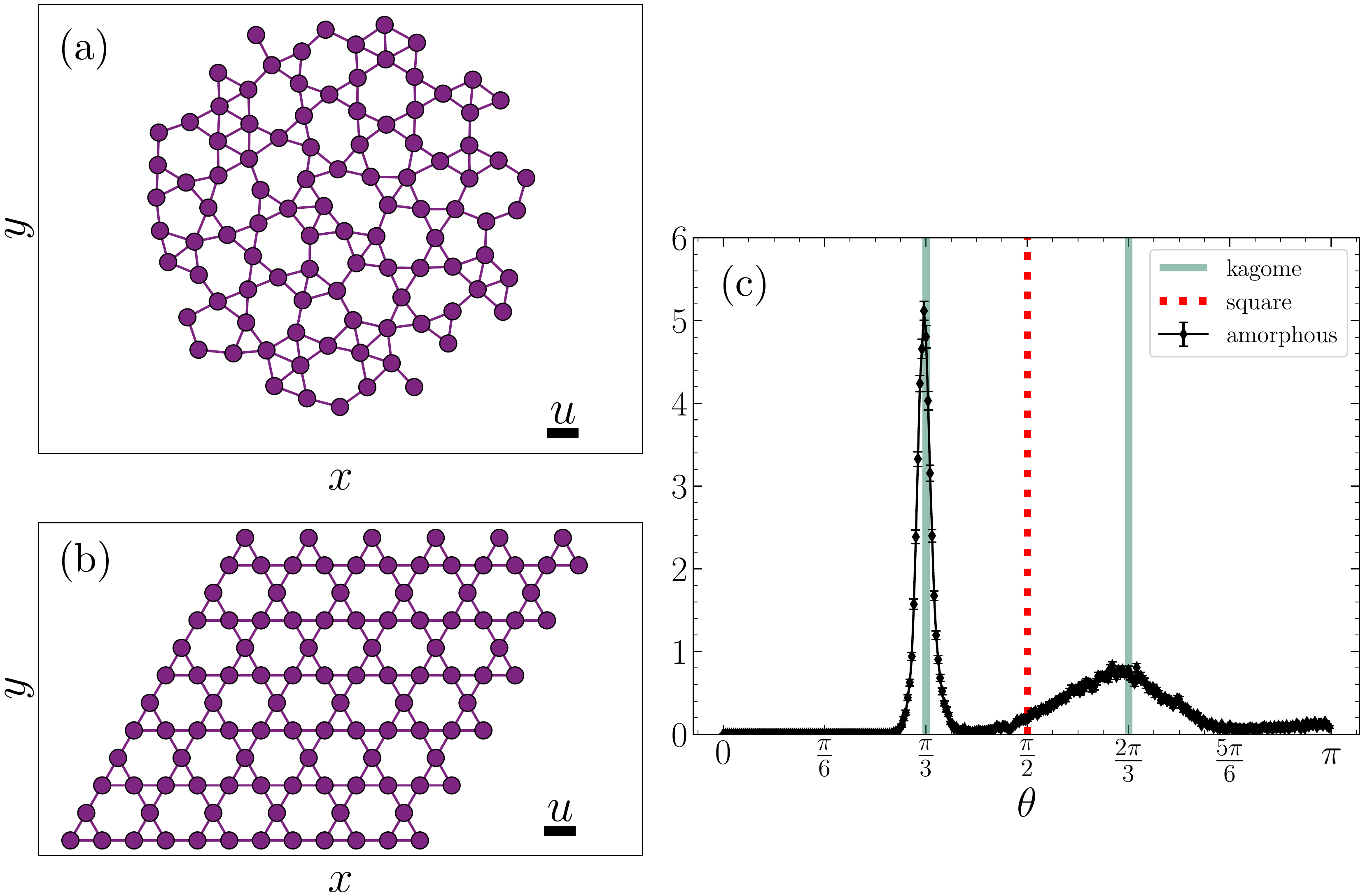}
\caption{Panels (a) and (b): representation of a 2D amorphous array (a) and of the kagome lattice (b).
Segments connect nearest-neighbor atoms.
Panel (c): probability distribution of bond-angle $\theta$ in the amorphous arrays. The vertical lines denote the angles of the kagome lattice (continuous lines) and of the square lattice (dashed line), with periodic boundary conditions.
}
\label{fig1}
\end{figure}

In this paper, we employ unbiased quantum Monte Carlo (QMC) simulations to investigate the ground-state properties of quantum Ising models defined on amorphous arrays. On the one hand, the coordination number and the preferred bond angles are set to mimic those of the periodic kagome lattice, as shown in Fig.~\ref{fig1}. On the other hand, the interactions are antiferromagnetic and decay with the sixth power of the distance, as in Rydberg-atom experiments. This setup is intended to combine the frustration effects occurring in the kagome lattice with positional disorder of the couplings. 
We determine the static magnetic structure factor and the replica-overlap Edwards-Anderson (EA) order parameter. The latter is suitable for identifying spin-glass phases, and we indeed find clear indications of a quantum phase transition from a paramagnetic to a spin-glass phase. The critical transverse field is determined via a finite-size scaling analysis. The magnetic structure factor indicates that, in the spin-glass regime, the antiferromagnetic correlations are short-ranged and isotropic, denoting the absence of spatial ordering.
We also analyze the spin-overlap distribution for the feasible sizes. To highlight the role of the amorphous structure, we make a comparison with results corresponding to the periodic kagome lattice.

Specifically, we simulate a 2D transverse field Ising Hamiltonian, which reads
\begin{equation}
    H = \sum_{i<j}J_{ij}\sigma_i^z\sigma_j^z - \Gamma\sum_{i=1}^N \sigma_i^x.
    \label{eq1}
\end{equation}

Here, $\sigma_i^{\alpha}$, for $\alpha=x,z$, are usual Pauli operators acting on the  spin $i=1,\dots,N$ at position $\mathbf{r}_i$. $\Gamma$ is the uniform transverse magnetic field, and the couplings $J_{ij}$ follow a power-low decay with respect to the distance: $J_{ij}=J_0/|\mathbf{r}_i-\mathbf{r}_j|^{6}$. $J_0$ is set as the energy unit.
These parameters describe the Ising model that can be implemented in neutral atom quantum simulators.
Notice that, to fulfill the $Z_2$ symmetry of the above Hamiltonian, local addressing of the detuning is required~\cite{PhysRevLett.132.263601,https://doi.org/10.1002/qute.202400291,PhysRevResearch.6.023031,kornjaca_2024_qrl}.

The amorphous configurations $\left(\mathbf{r}_1,\dots,\mathbf{r}_N\right)$ are created starting from the algorithm described in Ref.~\cite{juliafarre2024amorphous}. 
An exemplary array is visualized in Fig.~\ref{fig1}, panel (a). 
In short, the algorithm allows us to control the coordination number $C$, defined as the average number of atoms closer than the first minimum of the pair correlation function. 
Here, we set $C \approx 4$, which corresponds to either (clean) kagome or square lattices. In fact, an appropriate choice of the kernel function favors the former.  More details are provided in the Supplemental Material~\cite{SM}.
Indeed, the bond-angle distribution displays sizable broad peaks at the angles corresponding to the kagome lattice, namely, $\theta=\pi/3$ and $\theta=2\pi/3$, and essentially no weight at the angle $\theta=\pi/2$ occurring in the square lattice (see panel (c) of Fig~\ref{fig1}).
The average nearest-neighbor distance is $1.062(2)u$, where $u$ is the length unit shown in Fig.~\ref{fig1}.
To favor future studies, the amorphous arrays employed in this Article are provided via the public repository at Ref.~\cite{brodoloni_2025_15276169}.
To highlight the role of the lack of long-range structural order, below we also discuss the QMC results for the clean kagome lattice with periodic boundary conditions and nearest-neighbor distance $u$.

\begin{figure}[h]
\centering
\includegraphics[width=1.\columnwidth]{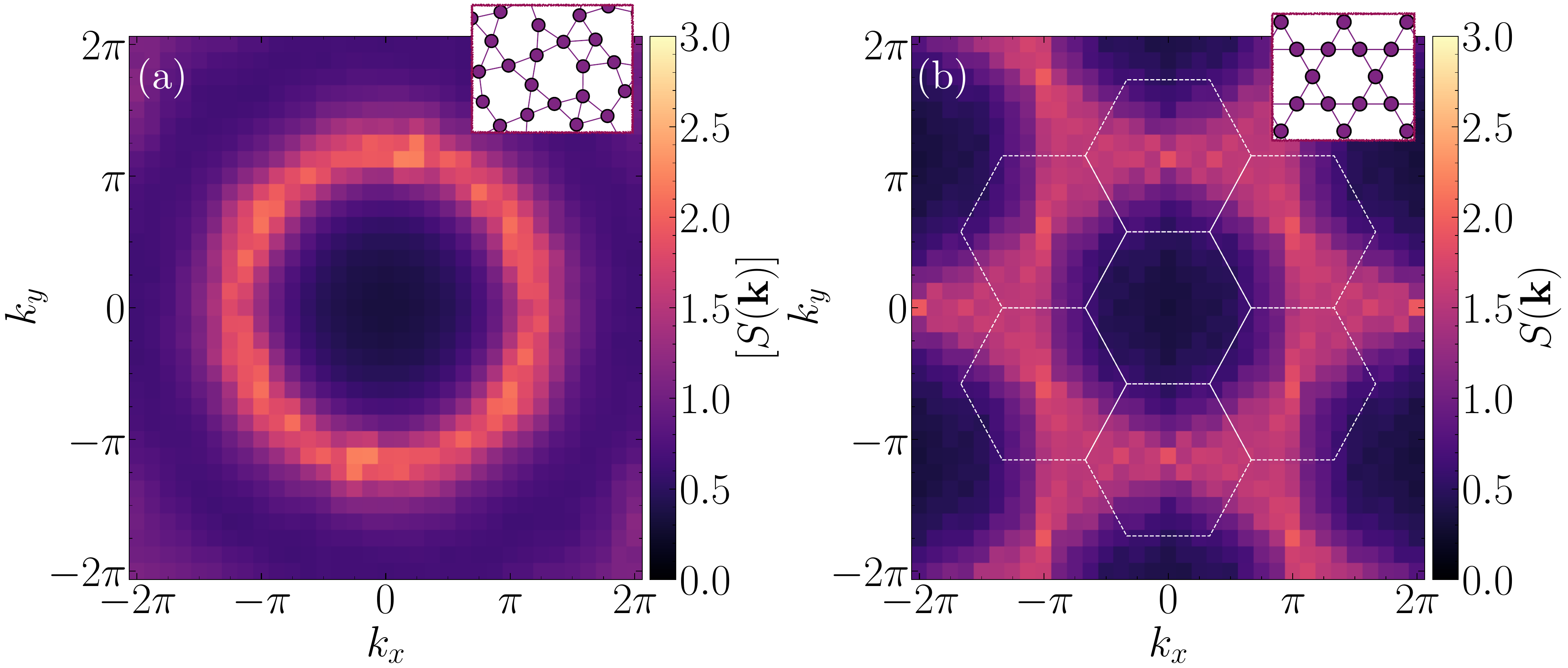}
\caption{Panels (a) and (b): magnetic structure factor $S(\mathbf{k})$ versus $k_x$ and $k_y$, with $\mathbf{k}=(k_x,k_y)$, for (ensemble-averaged) amorphous arrays (a) and for the periodic kagome lattice (b). In panel (b), the (white) hexagons represent the Brillouin zones.
}
\label{fig2}
\end{figure}

The ground state of the Hamiltonian~\eqref{eq1} is simulated using the continuous-time projection QMC algorithm detailed in Refs.~\cite{becca2017quantum,PhysRevB.98.235145,PhysRevE.101.063308}. 
It is verified that the possible systematic bias due to the population control~\cite{PhysRevA.97.032307,PhysRevB.103.155135,PhysRevB.105.235144} is below statistical uncertainties and the pure estimators are obtained via the standard forward-walking technique~\cite{PhysRevB.52.3654}. 
The guiding wave function is a neural network state in the form of a restricted Boltzmann machine~\cite{doi:10.1126/science.aag2302}, which is trained using the self-learning protocol of Ref.~\cite{PhysRevE.100.043301}. 
A relevant benefit of this QMC method is the absence of imaginary time discretization errors. 

On the other hand, finite-temperature path-integral Monte Carlo simulations with discrete imaginary time allow one to reach larger system sizes than those considered below~\cite{bernaschi2023quantum,BERNASCHI2024109101,king2023quantum}.
In general, QMC algorithms allow simulating power-law interactions without the truncations usually implemented in relevant alternative computational methods based on tensor-network methods~\cite{PhysRevB.78.035116}.
A viable QMC algorithm might also be the stochastic series expansion method~\cite{kramer2024quantumcritical,PhysRevLett.134.086501}.
For the amorphous arrays, open boundary simulations are performed, while for the kagome lattice we implement periodic boundary conditions and account also for the first periodic images beyond the fundamental simulation cell, given that the couplings with further-apart images are negligible.

\begin{figure}[h]
\centering
\includegraphics[width=1.\columnwidth]{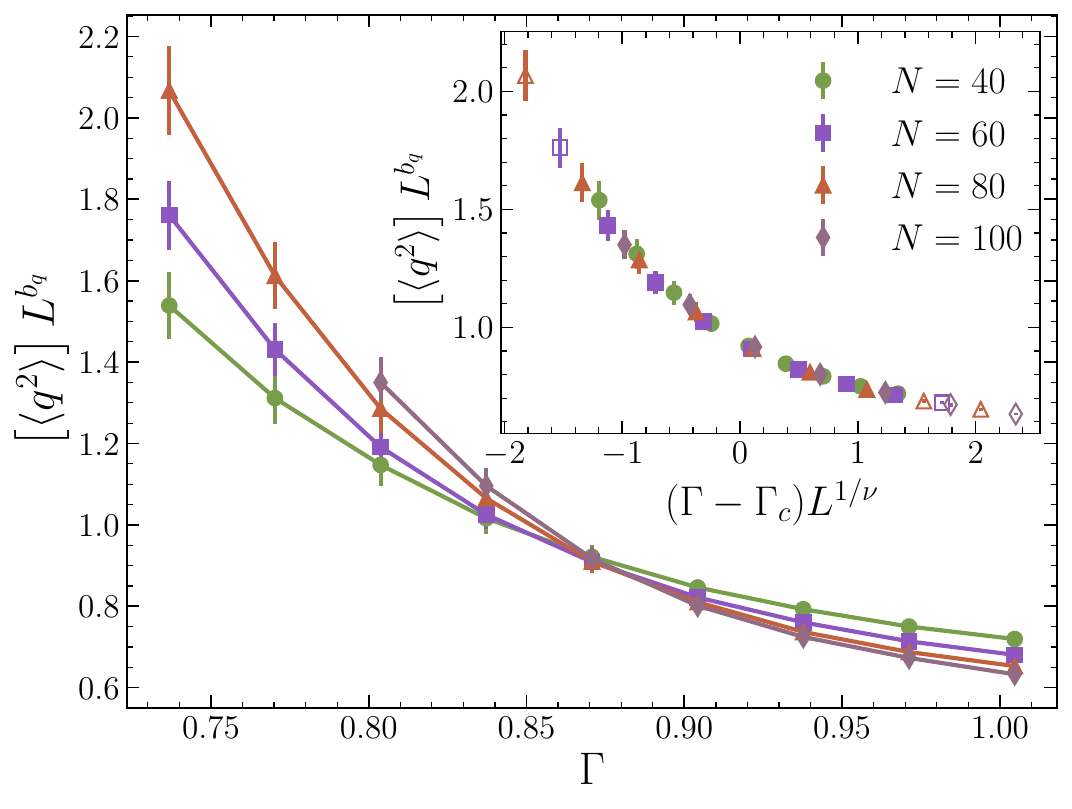}
\caption{Main panel: rescaled (disordered averaged) EA order parameter $\left[\left<q^2\right>\right]L^{b_q}$ as a function of the transverse field $\Gamma$. 
Different datasets correspond to different system sizes.
Inset: universal collapse of $\left[\left<q^2\right>\right]L^{b_q}$ plotted as a function of the transverse field $\left(\Gamma-\Gamma_c\right)L^{1/\nu}$. The critical parameters $\Gamma_c$, $1/\nu$, and $b_q$ are obtained by optimizing the data collapse~\cite{melchert2009autoscalepy}, excluding the data shown as empty symbols.
}
\label{fig3}
\end{figure}

The first observable we analyze is the static magnetic structure factor, defined as
$S(\mathbf{k})= \left< \mathcal{S}^z(\mathbf{k})\mathcal{S}^z(\mathbf{-k})\right>$,
where $\mathcal{S}^z(\mathbf{k})=N^{-1/2}\sum_i \sigma_i^z\exp\left(i \mathbf{k}\cdot \mathbf{r}_i\right)$.
For the amorphous arrays, the ensemble-average -- denoted with $\left[S(\mathbf{k})\right]$ -- is determined considering $N_r=30$ realizations.
The results at transverse field $\Gamma = 0.737$ display a broad circular hill, corresponding to isotropic magnetic correlations (see Fig.~\ref{fig2}). The peak height does not scale with the system size (see Supplemental Material~\cite{SM}), meaning that these correlations decay with distance. 
In the periodic kagome lattice, we again find a broad continuum, but this follows the hexagonal structure of the Brillouin zones. 
Notice that also in this case the correlations have a short-range character, pointing to a paramagnetic ground state, as found in previous studies addressing kagome Ising models with nearest-neighbor antiferromagnetic interactions~\cite{PhysRevLett.84.4457,PhysRevB.110.054432}. 
As discussed below, the system with $\Gamma=0.737$ is in the spin-glass regime.

To identify the spin-glass phase, we analyze the (mean squared) EA order parameter $\left< \hat{q}^2\right>$, which can be evaluated without systematic bias considering the overlap operator $\hat{q}$ between two identical replicas of the system~\cite{Young2006}; $\hat{q}$ is defined as
\begin{equation}
    \hat{q} = \frac{1}{N} \sum_{i=1}^N \sigma_{a,i}^z \sigma_{b,i}^z.
\end{equation}
Here, the subscripts $a$ and $b$ denote the two replicas, which are simulated within the projection QMC algorithm by evolving a two-fold replicated Hamiltonian $H_{T}=H_a+H_b$~\cite{PhysRevE.110.065305}, where $H_{a/b}$ are defined as in Eq.~\eqref{eq1} with Pauli matrices $\sigma_{a/b,i}^\alpha$.
The ensemble average $\left[\left< \hat{q}^2\right>\right]$ is performed over $N_r\in[100:200]$ amorphous arrays. 
Close to the critical point $\Gamma_c$, one assumes the following scaling Ansatz with a critical exponent $b_q$~\cite{king2023quantum}: $\left[\left<\hat{q}^2\right>\right]L^{b_q}=f\left(x\right)$, where $f(x)$ is a universal function of the reduced transverse field $x=(\Gamma-\Gamma_c) L^{1/\nu}$ and $\nu$ is the correlation-length critical exponent.
The linear size is defined as $L=\sqrt{N}$. The critical parameters are obtained by optimizing the data collapse using the software available from Refs.~\cite{melchert2009autoscalepy,pyfssa}, which minimizes deviations from a self-consistently derived master curve~\cite{PhysRevB.70.014418}.
Restricting the scaling analysis to $\left|x\right|\leq 1.34$, we obtain the following estimates:
$\Gamma_c = 0.864(33)$, $b_q=1.63(12)$, and $1/\nu=1.22(45)$.
Smaller fitting windows provide comparable estimates.
Interestingly, the universal critical parameters $b_q$ and $\nu$ are consistent with recent estimates for the 2D EA Hamiltonian on the square lattice with binary or Gaussian random couplings~\cite{king2023quantum,bernaschi2023quantum,PhysRevE.110.065305}, within the statistical errorbars.
The comparison is detailed in Table~\ref{table1}.
It is worth mentioning that the estimate of $1/\nu$ is expected to slightly decrease when very small sizes are excluded from the data collapse, due to corrections to the universal scaling ansatz~\cite{bernaschi2023quantum}. This effect cannot be discerned here due to the limited system sizes.
In fact, large-scale QMC simulations performed well within the spin-glass phase are  computationally demanding due to the slow stochastic dynamics.
Hence, we obtain a relatively large uncertainty on $1/\nu$, while the comparison of the parameter $b_q$ is more stringent.
As already noted~\cite{bernaschi2023quantum}, the values obtained for $b_q$ denote a very slow divergence of the spin-glass susceptibility $\chi_{\mathrm{SG}}=\left[\left<\hat{q}^2\right>\right]L^2$ at the critical point.
\begin{table}[]
\begin{tabular}{|c|c|c|c|c|}
\hline
Couplings       & $1/\nu$       & $b_q$        & $\Gamma_c$ & Ref.      \\ \hline
$50\%$ $\pm 1$  & $1.02(16)$    & $1.76(3)$    & $2.11(1)$  & \cite{king2023quantum}      \\
$50\%$ $\pm 1$  & $0.71(24)(9)$ & $1.73(8)(8)$ & $2.18(1)$  & \cite{bernaschi2023quantum} \\
$\mathcal{N}(0, 1)$       & $1.11(22)$    & $1.68(8)$    & $1.98(7)$  & \cite{PhysRevE.110.065305}    \\
$\propto 1/r^6$ & $1.22(45)$    & $1.63(12)$   & $0.864(33)$  & This work \\ \hline
\end{tabular}
\caption{Critical exponents $\nu$ and $b_q$ and transition point $\Gamma_c$ for different choices of random couplings $J_{ij}$, namely, binary random values ($50 \%$ $\pm 1$) and Gaussian values $\mathcal{N}(0, 1)$ on the square lattice, and the positional disorder decaying as $\propto 1/r^6$ considered in this Article. 
Numbers in parentheses denote the errorbar. Where two numbers are given, the first denotes the statistical uncertainty, the second the systematic effects.
}
\label{table1}
\end{table}

\begin{figure}[h]
\centering
\includegraphics[width=1.\columnwidth]{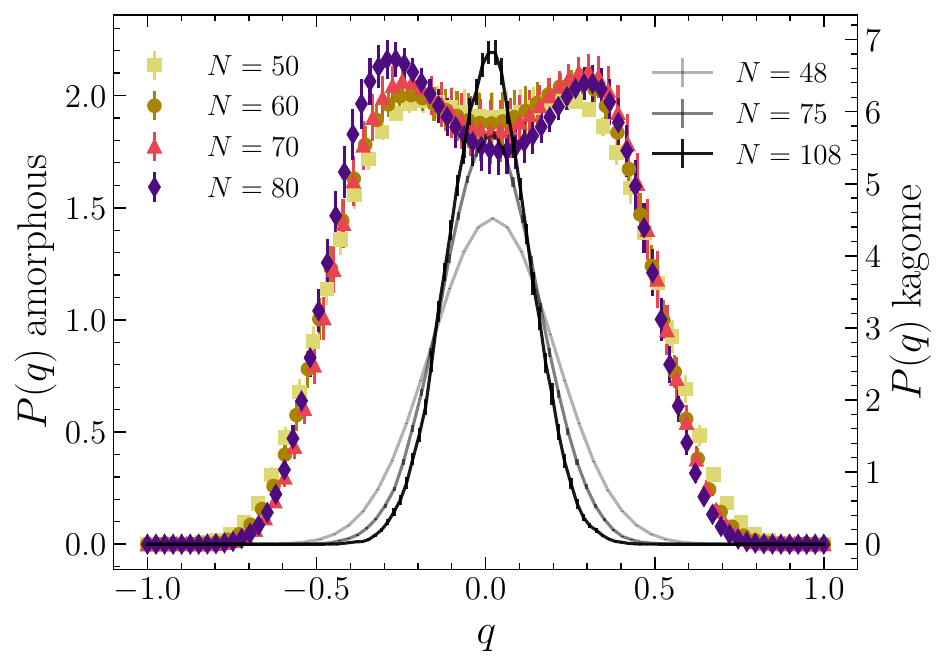}
\caption{Distribution of the replica spin-overlap distribution $P(q)$ for the 
(ensemble averaged) amorphous arrays (left vertical axis) and for the kagome lattice (right vertical axis). Different datasets correspond to different sizes $L$. The transverse field is $\Gamma=0.67<\Gamma_c$.
}
\label{fig4}
\end{figure}

To shed some light on the nature of the spin-glass phase of the amorphous arrays, we analyze the distribution of the ensemble-averaged replica spin-overlap $P(q)=\left[P_J(q)\right]$ at $\Gamma=0.67$; for a single realization, this distribution is computed as:
$P_J(q) = \left< \delta(q-\hat{q}) \right>$.
Simulations of $P(q)$ are often used to assess the applicability of any of the most prominent theories of the spin-glass phase, in particular the droplet-picture or the RSB scenario~\cite{marinari2000replica,PhysRevE.103.062111}.
While the first theory predicts two $\delta$-function peaks in the thermodynamic limit, the latter predicts a nontrivial distribution with sizable weight at $q=0$. 
Our QMC results at $\Gamma=0.67$ are shown in Fig.~\ref{fig4}. For the feasible sizes, we find a distribution with two (symmetry related) broad peaks and sizable values of $P(q=0)$. Yet, due to the limited system sizes and the statistical uncertainties, it is not possible to perform a reliable extrapolation of $P(q=0)$ to $L\rightarrow \infty$~\cite{PhysRevLett.81.4252,PhysRevB.62.946,PhysRevB.63.184422}. 
Furthermore, while for the smaller sizes essentially all realizations satisfy the expected symmetry $P_J(q)=P_J(-q)$, with the largest size $L=80$ some realizations break the symmetry for the feasible simulation time, leading to a small asymmetry of $P(q)$.
It is worth comparing the replica overlap distribution of amorphous arrays with the one corresponding to the periodic kagome lattice at the same transverse field. In the latter case, we find a narrow single-peak distribution that shrinks in the thermodynamic limit. 
Furthermore, the spin-glass susceptibility does not increase with system size, at least in the range $\Gamma\in[0.67,1.34]$~\cite{SM}. 
This highlights the important role of the absence of long-range structural order of the optical tweezers, compared with a periodic lattice with similar local structural properties.

We have investigated the ground-state properties of quantum Ising models with positional disorder using unbiased neural QMC simulations. The Hamiltonian was designed to describe Rydberg atoms arranged in 2D amorphous arrays, whose local structural properties mimic those of kagome lattices.
Notably, the finite-size scaling analysis of the EA order parameter revealed the occurrence of a quantum phase transition from a paramagnetic to a spin-glass phase. The critical exponents turned out to be consistent, within the statistical uncertainties, with those corresponding to the 2D EA model. 
By contrast, the results corresponding to the periodic kagome lattice display no hint of glassy behavior. This highlights the important role of the aperiodic structure and suggests that frustration effects in clean systems are not sufficient to induce a spin-glass phase.
Both the amorphous and the kagome setups host short-range antiferromagnetic correlations. However, these are isotropic in the former case, whereas they display a six-fold orientational symmetry in the latter. Notably, the ground state of the kagome lattice is paramagnetic in the explored parameter regime.

Our findings show that positionally distorted antiferromagnetic interactions can give rise to a spin-glass quantum phase transition, paving the way to the experimental observation of this phenomenon in Rydberg-atom platforms. 
As next steps, it would be interesting to explore if and how spin-glass behavior occurs in Hamiltonians without local detuning compensations and how the spin-glass phase could be prepared using quantum annealing protocols.
Temperature effects are also worth studying. It is known that, in the case of power-law decaying random interactions on periodic lattices, the classical spin-glass phase transition might occur at finite or zero temperature depending on the power~\cite{PhysRevLett.101.107203,PhysRevB.67.134410,berry2013monte}; on the other hand, the case of finite-temperature amorphous arrays has not been studied. Anyway, the 2D quantum phase transition has strong experimentally observable effects in the critical dynamics~\cite{king2023quantum}.
Paradigmatic spin-glass models can also be implemented in quantum annealers built with superconducting-flux qubits~\cite{king2023quantum}, but these devices do not allow readout at finite transverse fields, except indirectly via anneal-pause protocols~\cite{PhysRevB.110.054432}. 
Trapped ions are promising alternative platforms. Many-body localization has already been observed in a chain of ten ions~\cite{smith2016many}. Yet, ion traps are still limited to one-dimensional geometries and fewer spins, although small 2D crystals have recently been implemented~\cite{qiao2024tunable}. On the other hand, 2D arrays of several thousands of neutral atoms have already been realized~\cite{manetsch2024tweezer,PhysRevApplied.22.024073}.
Future studies could address other types of positional disorder, other forms of correlated aperiodic couplings, e.g., quasi-crystals featuring long-range orientational order, randomly distorted frustrated lattices, or longer-range interactions relevant for trapped ions~\cite{PhysRevA.108.012415}. 

\begin{acknowledgments}
    \emph{Acknowledgments}---
We thank L. Fallani, L. Guariento, V. Gavryusev, V. Vitale, and T. M. Santos for useful discussions.
Support from the following sources is acknowledged: 
PNRR MUR project PE0000023-NQSTI; 
PRIN 2022 MUR project ``Hybrid algorithms for quantum simulators'' -- 2022H77XB7; 
PRIN-PNRR 2022 MUR project ``UEFA'' -- P2022NMBAJ; 
National Centre for HPC, Big Data and Quantum Computing (ICSC), CN00000013 Spoke 7 -- Materials \& Molecular Sciences; 
CINECA awards ISCRA  IsCc2\_REASON, INF24\_lincoln, and INF24\_lincoln; 
EuroHPC Joint Undertaking for awarding access to the EuroHPC supercomputer LUMI, hosted by CSC (Finland), through EuroHPC Development and Regular Access calls. Pasqal's team acknowledges funding from the European Union the projects EQUALITY (Grant Agreement 101080142) and PASQuanS2.1 (HORIZON-CL4-2022-QUANTUM02-SGA, Grant Agreement 101113690)
\end{acknowledgments}

\emph{Data availability:}
To favor future numerical and experimental investigations, the coordinates of the amorphous configurations employed in this Article are made available at the repository in Ref.~\cite{brodoloni_2025_15276169}.
All other data are available from the corresponding author upon reasonable request.

\bibliography{bibliography}{}

\clearpage
\pagestyle{plain}
\onecolumngrid
\setcounter{page}{1}
\renewcommand{\thefigure}{S\arabic{figure}}
\setcounter{figure}{0}
\begin{center}
  {\large\textbf{Supplemental material for ``Spin-glass quantum phase transition in Rydberg atoms trapped in amorphous array''}}

  \vspace{2em}
  { L. Brodoloni$^{1,2}$ \orcidlink{0009-0002-0887-4020}, J. Vovrosh$^{3}$ \orcidlink{0000-0003-4034-5786}, S. Juli\`a-Farr\'e$^{3}$ \orcidlink{0000-0003-4034-5786}, A. Dauphin$^{3}$ \orcidlink{0000-0003-4996-2561}, S. Pilati$^{1, 2}$ \orcidlink{0000-0002-4845-6299}\\
  $^{1}$\small\textit{School of Science and Technology, Physics Division, Università di Camerino, 62032 Camerino, Italy}\\
  $^{2}$\small\textit{INFN, Sezione di Perugia, I-06123 Perugia, Italy}\\
  $^{3}$\small\textit{PASQAL SAS, 24 rue Emile Baudot - 91120 Palaiseau, Paris, France}\\}
\end{center}

\vspace{0.5em}
\begin{center}
\begin{minipage}{0.8\textwidth}
\hspace{10pt} This supplemental material provides additional details on: i) the generation of amorphous solids with controlled structural properties; ii) the extraction of independent patches from larger amorphous solids; iii) the dependence of the magnetic structure factor on the system size; iv) the spin-glass susceptibility and its scaling with the system size in the amorphous array and in the clean kagome lattice.
\end{minipage}
\end{center}

\section{SI. \space Amorphous solid generation}

The full procedure we use to generate the amorphous solid is detailed in Ref.~\cite{juliafarre2024amorphous}; here we give a brief overview of the finer details relevant to the results reported in this article. In particular, when generating the amorphous array, we target the coordination number $C=4$. $C$ can be understood as the average number of nearest-neighbor (NN) atoms or, more precisely, those closer than the first minimum of the pair correlation function from a reference atom. Furthermore, in this article, we tune the hyper-parameters of the loss function given in Ref.~\cite{juliafarre2024amorphous} to favor amorphous structures composed of kagome-type plaquettes rather than square-type ones, both of which have $C=4$. The core mechanism lies in the flexibility to control the NN separation, and the penalty for next-nearest neighbor (NNN) separations and beyond, both of which are governed by the choice of kernel function used in the design process. The NN separation is determined by the peak of the kernel; note, a sharp peak enforces uniform NN distances and hence low disorder, while a broader, flatter peak allows for greater disorder. Additionally, the rate at which the kernel decays away from the peak sets the penalty in the loss for atoms that are not NN. By tuning this decay rate, the system can be biased toward lattice types characterized by specific NN to NNN ratios; for example, $\sqrt{2}$ for square lattices or $\sqrt{3}$ for kagome lattices.

\section{SII. \space Amorphous patch extraction}
\begin{figure}[h]
\centering
\includegraphics[width=0.6\columnwidth]{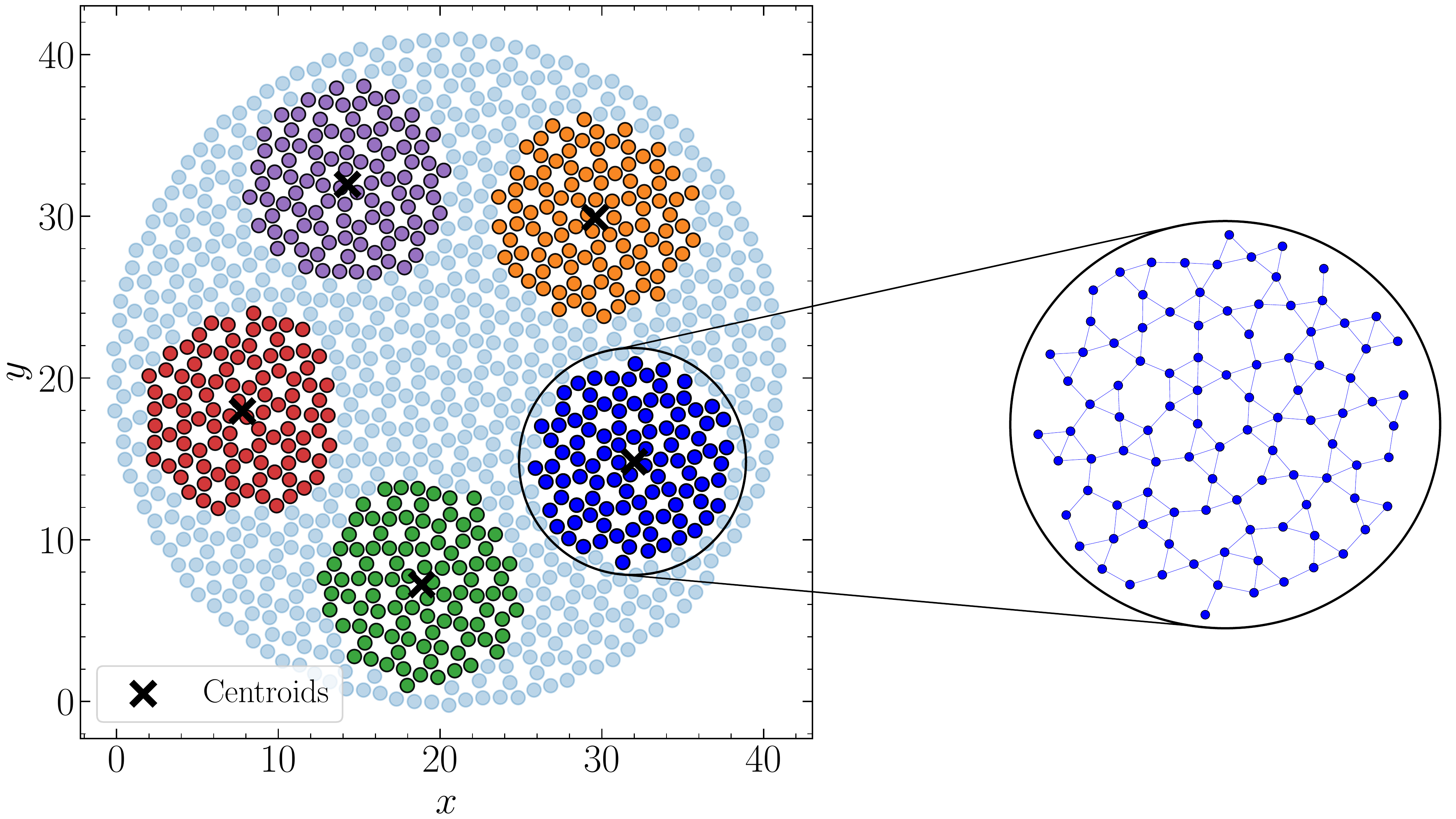}
\caption{Example of patch extraction. Points in light blue represent the entire amorphous array created with the algorithm described in Ref.~\cite{juliafarre2024amorphous}. The black crosses are the centroids identified by K-means clustering performed on the coordinates. The colored clusters are the five patches of $N=100$ atoms each. The zoom shows the detail of the blue patch.}
\label{figS1}
\end{figure}

In the study of spin glasses, all results have to be averaged over copious ensembles of different realizations of the random couplings. To optimize this generation process, we first create a few large amorphous arrays featuring $N\approx 1100$ atoms. We then extract from each array several well-separated patches, each containing $N\in[40:100]$ atoms. To avoid overlap among different patches originating from the same array, we apply K-means clustering on the atomic coordinates. The cluster centroids serve as reference points for the patch extraction, where we select the $N$ atoms closest to each centroid.
For each large array, we perform K-means clustering with five clusters. This number is chosen to ensure that each patch is independent, avoiding overlaps as well as boundary atoms, even for the largest patch size $N=100$.
A sketch of the extraction process is provided in Fig.~\ref{figS1}.
As discussed in the main text, the distribution of bond angles of the extracted patches exhibits broad peaks centered at the angles corresponding to the kagome lattice (with periodic boundary conditions), namely, $\theta=\pi/3$ and $\theta=2\pi/3$. A minimal weight is also observed at $\theta \approx \pi$, which we attribute to boundary atoms. On the other hand, the square lattice, which also corresponds to $C=4$, would lead to bond angles only at $\theta=\pi/2$.
The actual coordination number, averaged over bulk atoms of large arrays of size $N=1100$, is $C\simeq 4.01(3)$. Averaging over all atoms leads to a lower average $C\simeq 3.870(5)$ due to boundary effects.  When computed on the extracted patches, boundary atoms contribute more to the average; if these are taken into account, the coordination number is $C\simeq  3.57(2)$ for $N=100$.

\section{SIII. \space Magnetic structure factor for different system sizes}
\begin{figure}[h]
\centering
\includegraphics[width=0.6\columnwidth]{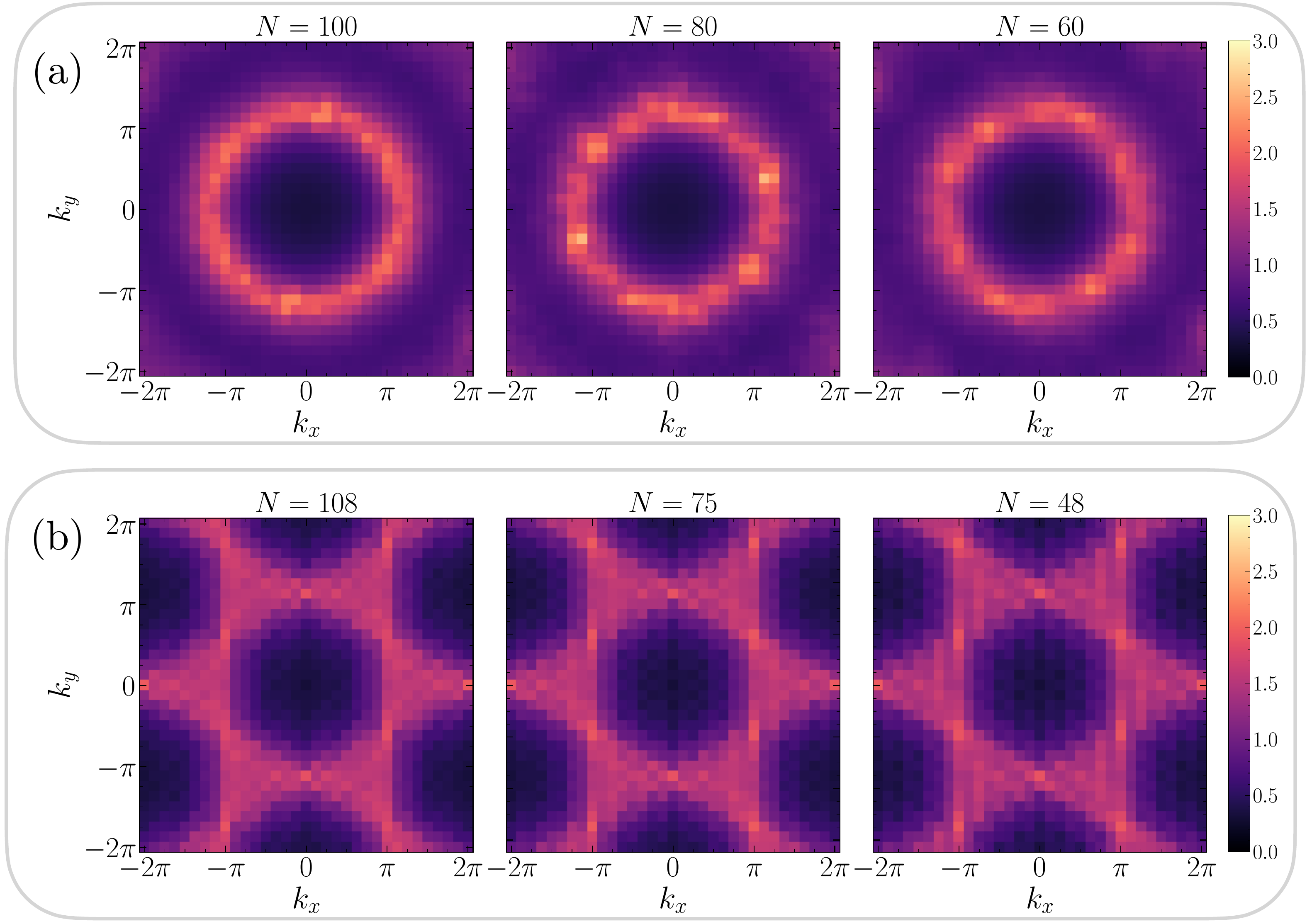}
\caption{Panel (a): Magnetic structure factor $S(\mathbf{k})$ for the amorphous patches of system sizes $N=[100, 80, 60]$ (from left to right). The results are averaged over $N_r=30$ realizations for $N=100$ and $N_r=10$ realizations for the other two sizes. Panel (b): Magnetic structure factor $S(\mathbf{k})$ for the clean kagome lattice for the system sizes $N=[108, 75, 48]$ (from left to right). }
\label{figS2}
\end{figure}

We compute the magnetic structure factor $S(\mathbf{k})$ (see definition in the main text) for amorphous patches of sizes $N=[100,80, 60]$. The ensemble-averaged structure factor, denoted as $[S(\mathbf{k})]$, is shown in Fig.~\ref{figS2}(a). It features very broad circular peaks, whose height does not significantly depend on the system size $N$. This indicates the lack of long-rage magnetic ordering.
Additionally, we compare this result with the structure factor of the periodic kagome lattice, shown in Fig.~\ref{figS2}(b). Also in this case, there are no sharp peaks, indicating short-range correlations. However, it is important to note that the pattern reveals the hexagonal structure of the Brillouin zones.
The results shown in Fig.~\ref{figS2} are obtained for a transverse field in the spin-glass phase of the amorphous system, namely $\Gamma=0.737$, for both the amorphous and periodic kagome systems. 

\section{SIV. \space Scaling of the spin-glass susceptibility}
\begin{figure}[H]
\centering
\includegraphics[width=0.6\columnwidth]{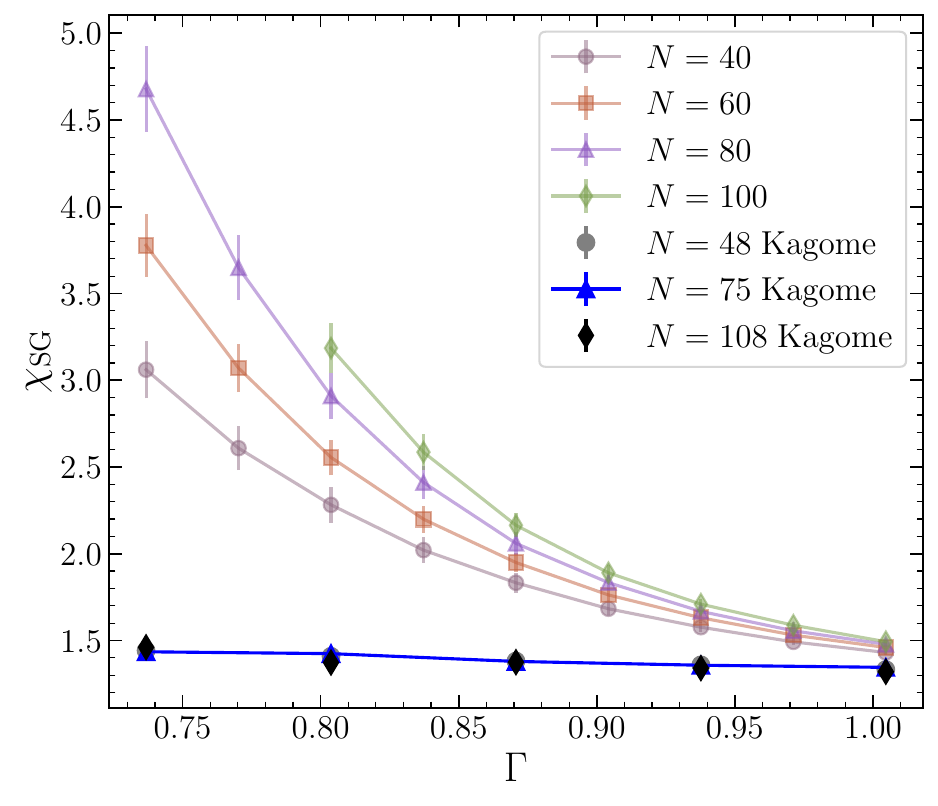}
\caption{Spin glass susceptibility $\chi_{\mathrm{SG}}$ for different system sizes $N$, for amorphous arrays (empty markers) and for the clean kagome lattice (filled markers)}
\label{figS3}
\end{figure}

Transitions to spin-glass phases can be signaled by a divergence of the spin-glass susceptibility $\chi_{\mathrm{SG}}$ with the system size $N$. In Fig.~\ref{figS3}, $\chi_{\mathrm{SG}}$ is plotted as a function of the transverse field $\Gamma$, both for (ensemble-averaged) amorphous arrays and for the periodic kagome lattice, for different system sizes. The difference between the two configurations is noticeable. In amorphous systems, $\chi_{\mathrm{SG}}$ clearly diverges for small $\Gamma$. On the other hand, the ordered kagome lattice  exhibits an essentially flat signal in the range $\Gamma\in[0.74,1.005]$, and no sizable dependence on the system size. This indicates the absence of long-range magnetic correlations, corresponding to a paramagnetic ground state. 

\end{document}